# Molecular distance matrix prediction based on graph convolutional networks

*Xiaohui Lin [a], Yongquan Jiang[*a, b, c] and Yan Yang [a, b, c]*

[a] School of Computing and Artificial Intelligence, Southwest Jiaotong University, Chengdu, 611756, China.

[b] Institute of Artificial Intelligence, Southwest Jiaotong University Chengdu, 611756, China.

[c] Manufacturing Industry Chains Collaboration and Information Support Technology Key Laboratory of Sichuan Province

[*]Email: yqjiang@swjtu.edu.cn

**Abstract**: Molecular structure has important applications in many fields. For example, some studies show that molecular spatial information can be used to achieve better prediction results when predicting molecular properties. However, traditional molecular geometry calculations, such as density functional theory (DFT), are time-consuming. In view of this, we propose a model based on graph convolutional networks to predict the pairwise distance between atoms, also called distance matrix prediction of the molecule(DMGCN). In order to indicate the effect of DMGCN model, the model is compared with the model DeeperGCN-DAGNN and the method of calculating molecular conformation in RDKit. Results show that the MAE of DMGCN is smaller than DeeperGCN-DAGNN and RDKit. In addition, the distances predicted by the DMGCN model and the distances calculated by the QM9 dataset are used to predict the molecular properties, thus showing the effectiveness of the distance predicted by the DMGCN model.

**Keywords**: Molecular structure; Graph convolutional neural network; Distance matrix.

# 1 Introduction

In this era of rapid development of artificial intelligence, artificial intelligence technology, especially machine learning methods, has become an important part of the technology industry. At the same time, it has also been applied in chemical, biological, material discovery and other professional fields [1]. Machine learning methods are mostly used to predict the properties of molecules in chemistry, biology and material discovery. When using traditional machine learning methods to predict the properties of substances, some known properties are often used to predict certain properties, and these features are usually manually selected [1,2,3]. Later, because the properties of matter are closely related to its molecular structure, more people choose to use graph neural network to predict its properties according to the molecular graph or molecular spatial information, and the molecular spatial information helps to improve the prediction results [4,5,6,7,8,9], so it is necessary to know the spatial information of molecules in advance. Spatial information is a representation of molecular geometry. In many theoretical experiments of chemical simulation, the structure of substances also plays a very important role. Therefore, predicting the spatial information of new substances that do not exist is very useful for better predicting their properties and downstream tasks such as their combination, which helps to accelerate the discovery of new substances.

At present, the traditional methods to obtain molecular geometry are usually determined by experimental methods or calculated by molecular simulation. The experimental methods include microwave spectrum, X-ray diffraction, electron diffraction and neutron diffraction. The electron diffraction of gas sprays gas molecules into the diffraction cavity, injects accelerated electrons and measures the diffraction data. By measuring the diffraction data, the distance matrix of molecule is obtained, which is also equivalent to obtaining the molecular structure. However, the experimental methods are often complex, and the experimental equipment like neutron diffraction still needs high cost. The traditional molecular simulation calculation methods mainly use density functional theory(DFT) for calculation and optimization, and these calculation methods have many calculation steps and high calculation cost measured in hours. In addition, although some empirical calculation methods have high calculation efficiency, the calculation results are often poor. In recent years, some studies reconstruct atomic coordinates by predicting the distance matrix to obtain the geometric structure of molecules [10,11,12,13,14, 23]. Most of these works use probabilistic generative model and aim to generating multiple low-energy conformations, but there are still some limitations. For example, [13] can only deal with molecules composed of fixed atoms, and [10, 12] need to obtain the distance between atoms through one conformation first. In view of this situation, this paper proposes a molecular distance matrix prediction model (DMGCN) based on graph convolution network [15]. The model takes the graph representation converted from the simplified molecular input line entry system (SMILES) of molecules as the input to predict the pairwise distance between atoms in molecules, which is used for downstream tasks such as reconstruction of molecular geometry or prediction of molecular properties.

The main contributions of this paper are as follows: (1) A new graph convolutional network model for molecular distance matrix prediction is proposed; (2) A comparative experiment is conducted to explore and analyze the influence of different data sets with different maximum atomic numbers on our model, and the results show that our model has a smaller MAE than DeeperGCN-DAGNN and RDKit, and the bond length predicted by our model is close to that calculated by DFT and measured by experiment; (3) Using the predicted distance matrix to predict the molecular properties, to reveal the effectiveness of our method.

## 2 Method

### 2.1 Encoding

In this paper, the molecule is defined as a graph G = (V, E), where the graph node V represents the atom and the edge E represents the bond. The feature vector of the atom is defined as v, v∈V, and the feature vector of the bond is defined as e, e∈ E, where v is encoded as a 19-dimensional feature vector and e as a 4-dimensional feature vector. The specific vector composition is shown in Table 1.

**Table 1. Feature vector composition**

|      | Vector name | Encoding description |
|------|-------------|----------------------|
| **atom** | molecular composition | 5 - dimensional vectors, the number of atoms of each element in a molecule. |
|      | atomic type | 5 - dimensional vector, one-hot coding of CNOFH |
|      | atomic number | |
|      | covalent radius | |
|      | electron arrangement | 2 - dimensional vector, CNOFH have at most two electron layers |
|      | electronegativity | |
|      | first ionization energy | the Log transformation |
|      | atomic radius | |
|      | atomic orbital | 2 - dimensional vector |
| **bond** | bond type | 5 - dimensional vector, one-hot coding of single bond, double bond, triple bond, aromatic bond and no bond |

### 2.2 Model

The DMGCN model proposed in this paper takes the SMILES as the input which is transformed into the graph in the preprocessing stage, and the overall frame diagram of the model is shown in Figure 1. Due to the need to predict the molecular distance matrix, so when building the input graph, connecting the atoms which don't have chemistry bond between them is necessary, namely to construct a complete graph. But complete graph destroyed the original structure of molecular graph information, therefore, the model is set up two branches to process the complete graph and the molecular graph, respectively. The branch where the molecular graph is located needs to accomplish two things: one is to predict the distance between bonded atoms, and the other is to send the molecular graph into the update layer with graph convolution operation [15] for node updating. After the update, the feature vectors of all nodes are assigned to the nodes in the complete graph, which uses the updated nodes to update the edge between atoms without bonds. After going through 3 update layers, the corresponding edges after each updating layer are concatenated to predict the distance of atoms without bonds.

A multi-layer perceptron (MLP) containing 3 hidden layers is used to predict the distance of atoms [16], which is defined as follows

$$e_i = \varphi(W_i * e_{(i-1)}) + b_i , \quad (1)$$

where $e_i$ represents the output vector at the $i$-th layer, $W$ is the trainable weight parameter, $b$ is the bias, and $\varphi$ is the activation function, which makes the function non-linear for better approximating the real value. ReLU is used as the activation function here.

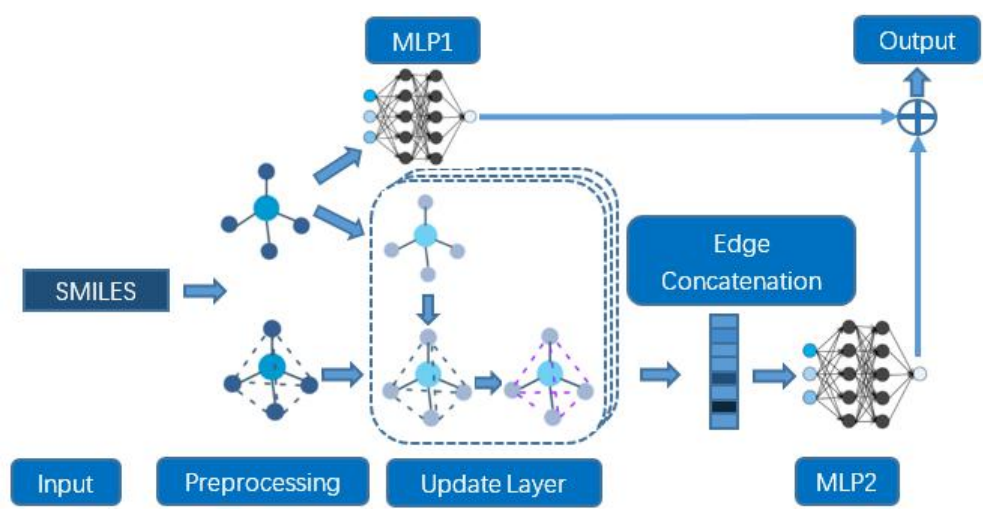

Figure 1. Model Structure. (1) Preprocessing: Constructing molecular graph and complete graph according to the input SMILES; (2) Update Layer: The nodes are updated by graph convolution on the molecular graph and the edges in the complete graph are updated according to the updated nodes. (3) Edge Concatenation: Concatenating the edge feature vectors in each update layer and the nodes feature vectors in the last update layer; (4) MLP1 takes the concatenation of the edge feature vectors and the corresponding nodes feature vectors in molecular graph as the input and MPL2 takes the result of Edge Concatenation as the input.

## 2.3 Update layer

In the update layer, the main purpose is to update the feature vectors of nodes and edges. The main operation is graph convolution operation. For each node in the molecular graph, its neighbor nodes are aggregated to update itself. The aggregation operation is shown in Formula (2)

$$h_{N(i)}^{(l+1)} = aggregate(\{h_j^l, \forall j \in N(i)\}) , \qquad (2)$$

where N(i) represents all the neighbor nodes of node *i*, l represents the number of graph convolution layers, and aggregate represents the aggregation method. In this paper, the average value is used as the aggregation method. After the aggregation information of neighbors is obtained, it is concatenated together with node *i*, sent into the full connection layer, and the updated node features of node *i* are output. The node update formula is shown in Formula (3)

$$h_i^{(l+1)} = \varphi(f(h_i^l || h_{N(i)}^{(l+1)})) , \qquad (3)$$

where || represents the operation of concatenation, $\varphi$ means activation function ReLU, $f$ means full connection layer. When all nodes in the molecular graph are updated, new node features are assigned to nodes in the complete graph, and then edges between atoms without bonds in the complete graph are updated. The operation is shown in Formula (4)

$$e_i = \varphi(f(v_i^r || v_i^s || e_{(i-1)})) , \qquad (4)$$

where $e_i$ represents the feature vectors of the *i*-th edge, $v_i^r$ and $v_i^s$ represent the feature vectors of the two nodes forming the edge. When all edges are updated, the update layer ends. After passing through 3 such updating layers, the edge feature vectors obtained in each update layer and the node features forming the edge in the last layer are concatenated together and sent into the multi-layer perceptron to predict the length of the edge. The operation is shown in Formula (5)

$$e = \varphi(f(e^0||e^1||\ldots||v_r||\,v_s))\,, \tag{5}$$

where $e$ represents the predicted distance between all atoms without bonds, $e^i$ means the edge features after $i$-th update layer, $v_r$ and $v_s$ represent the feature vectors of the two nodes that constitute each edge in the last update layer.

It is worth mentioning that after the convolution operation of each update layer, the batch normalization layer (BN layer) is used to prevent overfitting and gradient disappearance.

## 2.4 Model constructing and training

The model is built based on PyTorch framework. In order to achieve rapid convergence of the model, multiple molecular maps are formed into a batch for training. The batch size is 2048, and a total of 200 epochs are trained. Adam optimizer is used to optimize gradient descent in the model, and the learning rate is 0.01. In the later stage, the value of learning rate would be dynamically adjusted in order to achieve better convergence. The total loss function of the whole model is defined as Formula (6):

$$Loss = \alpha loss_1 + \beta loss_2\,, \tag{6}$$

where $loss_1$ represents the loss of MLP1 and $loss_2$ represents the loss of MLP2. Through the experiment, α and β are set as 0.4 and 0.6 respectively, and the mean absolute error is used for both of these losses, which are defined as

$$loss = \frac{1}{n}\sum_{i=1}^{n}|y_i^{label} - y_i^{predict}|\,. \tag{7}$$

where n is the number of edges in each batch, $y^{label}$ represents the label, and $y^{predict}$ represents the predicted value calculated by the model.

The model is trained on Intel(R)Core(TM) I7-4710MQ CPU. Figure 2 shows the decline curve of the loss function when using a data set less than 11 atoms for training. The DMGCN model can converge within 100 epochs, while with the increase of the number of samples, the model can converge within fewer epochs.

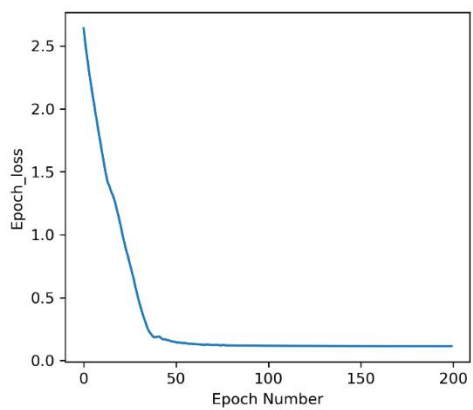

Figure 2. Loss decline curve.

# 3 Result

## 3.1 Dataset

In this paper, we use the QM9 dataset to train and test our proposed model. The molecules in the dataset consist of elements such as C, N, O, F, H and contain up to 9 heavy atoms. The atomic coordinates in the data set are first generated by the initial Cartesian coordinates of Corina's parse molecule Smiles. Then MOPAC is used to perform geometric relaxation at the semi-empirical theoretical level of PM7, and the results are used as the input to the geometric relaxation of Gauss B3LYP [17] to obtain the optimized atomic coordinates. There are 133885 molecules in the dataset. After removing the molecules containing incorrect data which contains characters that cannot be processed, the remaining 131,808 molecules are used as the data set of this paper.

## 3.2 Comparative experiments

In order to study the influence of atomic number difference in the training set on the model, the subsets composed of molecular data whose atomic number is less than 11 and 15 are screened out from the QM9 dataset and all the data are taken as the data set to train the model. We divide the dataset according to the proportion of 9:1 between the training set and the test set, and manually select the hyper-parameter on the subset with less than 11 atoms. In addition, in order to illustrate the advanced nature of the method, our model is compared with the method used in the RDKit chemical calculation package to calculate the geometric structure of atoms [18] and the DeeperGCN-DAGNN model in [23]. As far as we know, DeeperGCN-DAGNN is the best model of the same kind. These three methods are compared in the same environment and on the same dataset. The results are shown in Table 2.

Methods in RDkit are mainly based on the distance of the method and the method based on knowledge (ETKDG) The distance-based method is to generate the optimized molecular boundary matrix, randomly generate the distance matrix according to the boundary matrix, and then map the distance matrix to the three-dimensional space to generate atomic coordinates. Finally, the force field is used to roughly optimize the atomic coordinates [19]. The molecular structure generated by this method is relatively rough, and it needs to be optimized by force field again, while the knowledge-based method is based on some rules summarized by Riniker et al. from the small molecular structure of crystal structure database to modify the results obtained by the distance-based method [20]. The average time calculated by the three methods on the test set with the maximum number of atoms of 11 is shown in Table 3.

**Table 2.** The test results of RDKit, DeeperGCN-DAGNN and DMGCN (Unit: Å)

| Maximum number of atoms | Size of dataset | Method | MAE | RMSE |
|---|---|---|---|---|
| **11** | 2075 | RDKit | 0.1466 | 0.2433 |
|  |  | DeeperGCN-DAGNN | 0.3200 | 0.4520 |
|  |  | Ours | **0.0906** | **0.2131** |
| **15** | 25706 | RDKit | 0.2340 | 0.3960 |
|  |  | DeeperGCN-DAGNN | 0.2160 | **0.3260** |
|  |  | Ours | **0.1350** | 0.3400 |
| **all** | 131808 | RDKit | 0.3480 | 0.5630 |
|  |  | DeeperGCN-DAGNN | 0.2980 | **0.4380** |
|  |  | Ours | **0.1697** | 0.4471 |

**Table 3.** The average time calculated by the three methods on the test set with the maximum number of atoms of 11. (Unit: Second)

| Method | Average time |
|---|---|
| **RDKit** | 0.0148 |
| **DeeperGCN-DAGNN** | 0.0125 |
| **ours** | 0.0138 |

As you can see from Table 2, with the increasing difference of the number of atoms in the dataset, whether it is method in RDKit or our model, the errors generated are gradually increasing. This is because as the number of atoms increases, the distance differences between atoms without bonds also increase, making the model more difficult to fit. However, it can be seen that our error is generally smaller than RDKit method, and with the increasing difference in the number of atoms, the error growth range of RDKit is larger than that of our model, which also indicates that our model has better robustness compared with RDKit. Besides, the DeeperGCN-DAGNN performs poorly when the amount of data is small, but when the amount of data is sufficient, the MAE is between ours and RDKit, and the RMSE is slightly higher than our model. Overall, our model achieved good results.

To show the effect of the model more visually, the predicted results of interatomic distance with the error around the mean error are selected from the test set of less than 11 atoms and compared with the ground truth. As shown in the Figure 3, the distances between atoms with bonds (Figure 3.a) are relatively small and approach the ground truths, while the distance errors between atoms without bonds (Figure 3.b) are relatively large, but the trend is basically consistent with the ground truth.

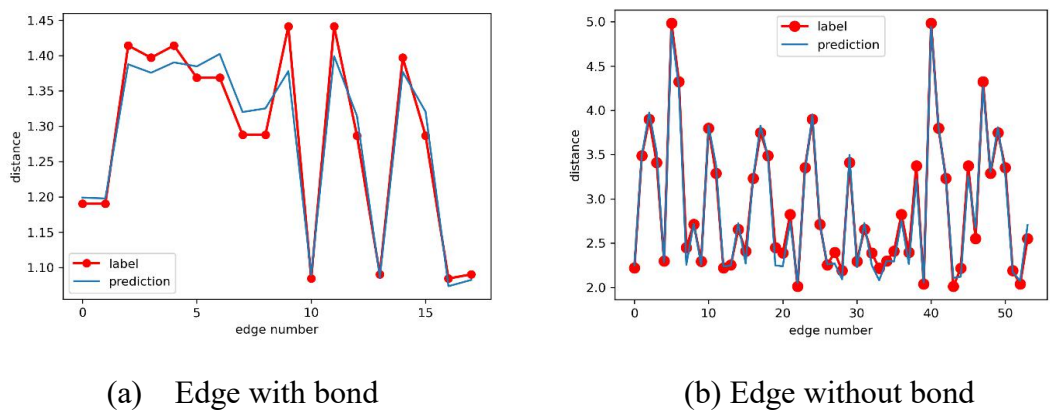

(a) Edge with bond    (b) Edge without bond

Figure 3. Prediction of interatomic distance for a randomly selected molecule, the unit of distance is Å, and the x-axis represents the number of edges, and the order of edges is determined by their order in the distance matrix; (a) the distance between atoms with bonds; (b) the distance between atoms without bonds.

In addition, in order to facilitate the observation of the distance prediction error distribution between atoms with bonds, we calculate the absolute error of the distances between atoms with bonds for the molecules whose atom number is less than 15. The statistical results are shown in Figure 4. Most errors are concentrated between 0 and 0.02 Å, with a MAE of 0.0208 Å and a RMSE of 0.0301 Å. It can be seen that the model is more accurate in predicting the distances between bonded atoms.

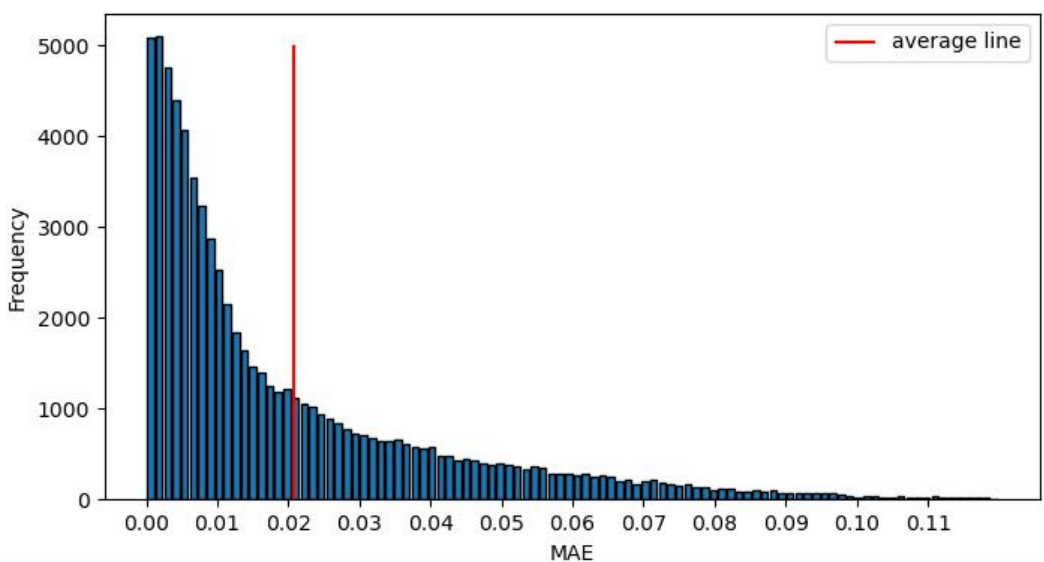

**Figure 4**. Distance prediction error distribution between bonded atoms. The red line represents the average of the errors.

Furthermore, Table 4 shows the errors among the bond lengths predicted by DMGCN, the bond lengths calculated by B3LYP/6-31G (2df, p) and experimental bond lengths on some molecules which come from the intersection of set B in [21] and QM9 dataset. In table 4, Error-1 represents the absolute value of the difference between the bond lengths predicted by the model and the bond lengths calculated by B3LYP/6-31G (2df, p) method (the method is used to calculate the molecular spatial structure in QM9 dataset), Error-2 represents the absolute value of the difference between the bond lengths calculated by B3LYP/6-31G (2df, p) method and the bond lengths obtained by experiment, and Error-3 is the absolute value of the difference between the bond lengths predicted by the model and the bond lengths obtained by experiment. Among the three errors, the one with the least error uses bold font, followed by green font. It can be seen from Table 4 that the bond length obtained by B3LYP/6-31G (2df, p) method is extremely close to the experimental value, besides there are more small values in Error-3 than in Error-1, which indicates that there are more bond lengths predicted by DMGCN closer to the experimental value than to the value calculated by B3LYP/6-31G (df, p) method.

**Table 4**. Pairwise error among the bond lengths predicted by DMGCN, bond lengths calculated by B3LYP/6-31G(2df, p) and experimental bond lengths. (Unit: Å)

| Molecule | Bond | Our Prediction | B3LYP/6-31G(2df, p) | Experiment | Error-1 | Error-2 | Error-3 |
|---|---|---|---|---|---|---|---|
| $CH_4$ | CH(1) | 1.0976 | 1.0920 | 1.085 | **0.0056** | 0.0070 | 0.0126 |
|  | CH(2) | 1.0976 | 1.0920 | 1.085 | **0.0056** | 0.0070 | 0.0126 |
|  | CH(3) | 1.0976 | 1.0919 | 1.085 | **0.0056** | 0.0069 | 0.0126 |
|  | CH(4) | 1.0976 | 1.0919 | 1.085 | **0.0056** | 0.0069 | 0.0126 |
| $NH_3$ | NH(1) | 1.0030 | 1.0172 | 1.012 | 0.0142 | **0.0052** | 0.0090 |
|  | NH(2) | 1.0030 | 1.0172 | 1.012 | 0.0142 | **0.0052** | 0.0090 |
|  | NH(3) | 1.0030 | 1.0172 | 1.012 | 0.0142 | **0.0052** | 0.0090 |
| H2O | OH(1) | 0.9439 | 0.9621 | 0.957 | 0.0182 | **0.0051** | 0.0131 |
|  | OH(2) | 0. 9439 | 0.9621 | 0.957 | 0.0182 | **0.0051** | 0.0131 |
| C2H2 | C(1)C(2) | 1.2063 | 1.1991 | 1.203 | 0.0072 | **0.0039** | 0.0033 |
|  | C(1)H(1) | 1.0863 | 1.0621 | 1.061 | 0.0242 | **0.0011** | 0.0253 |

|  | C(2)H(2) | 1.0863 | 1.0621 | 1.061 | 0.0242 | **0.0011** | 0.0253 |
| --- | --- | --- | --- | --- | --- | --- | --- |
| **HCN** | CN | 1.1573 | 1.1517 | 1.153 | 0.0056 | **0.0013** | 0.0043 |
|  | CH | 1.0843 | 1.0666 | 1.065 | 0.0177 | **0.0016** | 0.0193 |
|  | CO | 1.2067 | 1.2000 | 1.208 | 0.0067 | 0.0080 | **0.0013** |
| **H2CO** | CH(1) | 1.0988 | 1.1109 | 1.116 | 0.0121 | **0.0051** | 0.0172 |
|  | CH(2) | 1.0988 | 1.1109 | 1.116 | 0.0121 | **0.0051** | 0.0172 |

Error-1: The absolute value of the difference between Prediction and B3LYP/6-31G(2df, p);

Error-2: The absolute value of the difference between B3LYP/6-31G(2df, p) and Experiment;

Error-3: The absolute value of the difference between Prediction and Experiment;

### 3.3 Ablation experiments

We use BN layer and double branch structure in the design of the model. In addition, for the sake of improving the prediction accuracy of the model, we add global information (molecular composition) to the node features, the edge features use five-dimensional features and the edge features between atoms without bonds are the sum of edges on the shortest path between two atoms. In order to show the effectiveness of these parts of the model, we perform ablation experiments on dataset with the number of the molecular atoms less than 15. The results are shown in the table 5. It can be seen that each part mentioned above, especially double branch and edge feature with shortest path, has an important impact on the model.

**Table 5**. The results of ablation experiments in model (Unit: Å)

| Model | MAE | RMSE |
| --- | --- | --- |
| **Model with double branch** | **0.1350** | **0.3400** |
| Model with single branch | 0.2499 | 0.3748 |
| Model without BN | 0.1497 | 0.3655 |
| **Edge feature without shortest path** | 0.1975 | 0.4764 |
| **Node feature without MC** | 0.1528 | 0.3680 |

BN: batch normalization layer; MC: molecular components;

## 3.4 Comparison of property prediction

In order to indicate the effectiveness of the proposed method, we use the method in [22] to predict the properties of the molecule respectively according to the distance matrix of molecule calculated from the QM9 and the distance matrix predicted by our method. We divide the whole data set into training sets and test sets in 9:1 proportions, and train the model with the training set. Then we use the trained model to calculate the molecular distance matrix in the test set, and then divide the data in the test set into training set and test set in 8:2 ratios for training the model in [22] to predict the properties of the molecules, such as the highest occupied molecular orbital energy ($\epsilon$HOMO) and the lowest non-occupied molecular orbital energy ($\epsilon$LUMO).

The results are shown in the Table 6. It can be seen that the predicted results of properties based on molecular distance matrix predicted by our model (Predicted) are close to that based on molecular distance matrix calculated by the data in QM9. This indicates that the intermolecular atomic distance predicted by DMGCN can be applied to reality and the error is within the acceptable range.

**Table 6**. Error comparison of property prediction using atomic distance in QM9 and predicted atomic distance (Unit: eV)

| Properties | MAE | | Difference |
| --- | --- | --- | --- |
| | QM9 | Predicted | |
| $\epsilon$HOMO | 0.1619 | 0.1645 | 0.0026 |
| $\epsilon$LUMO | 0.1901 | 0.1736 | 0.0165 |

## 4 Conclusions

From what has been discussed above, this paper proposes a two-branch DMGCN model based on graph convolutional network to solve the problems of high experimental cost and high cost of traditional calculation methods in determining molecular structures. The model takes the molecular SMILES expression as the input and converts it into graph representation through data pre-processing, which is used to predict the distance between any two atoms in the molecule to achieve the purpose of determining the molecular structure. In order to show the advance of the method, the prediction results of this model is compared with DeeperGCN-DAGNN model and method in RDKit chemical calculation package, and obtains better MAE. Results also show that the bond length predicted by our model is close to that calculated by DFT and measured by experiment. In addition, as the atomic number difference between molecules in the dataset increases, the error of our model has a smaller increase than that of RDKit, showing better robustness. Subsequently, in order to reveal the effectiveness of our proposed model, experiments on molecular property prediction are also conducted. The results showed that the error of molecular property prediction based on the distances predicted by our model is close to that based on the calculated distance of QM9, and the error difference between the two is within an acceptable range. Although the prediction results of DMGCN are generally better than RDKit, there's a lot of improvement space in the future. In the next step, we will consider how to let the model learn more overall features of molecules, so as to reduce the prediction error of the distance between atoms without bonds, and get a more accurate molecular distance matrix.

## Acknowledgements

The research work is supported by the National Natural Science Foundation of China (No.61976247).

## Notes

The authors declare no competing financial interest.